\documentclass[twocolumn,showpacs,preprintnumbers,amsmath,amssymb,%
prd,a4paper,floatfix,nofootinbib,superscriptaddress]{revtex4-1}  

\usepackage{amssymb,amsfonts,amsmath,bm,dsfont}
\usepackage{graphicx}
\usepackage{wrapfig}
\usepackage{caption,subcaption}
\usepackage{multirow}
\usepackage{epsfig}
\usepackage{epsf}
\usepackage{color} 
\makeatletter
\usepackage{float}
\renewcommand{\@makefntext}[1]{\parindent=1em\noindent\hbox to 1.8em{\hss$^{\@thefnmark}$}#1}
\renewcommand{\@footnotemark}{\hbox{\mathsurround=0pt$^{\@thefnmark}$}}

\makeatother

\usepackage[tight]{units}

\newcommand{\fig}[1]{Fig.~{\ref{#1}}}
\newcommand{\tab}[1]{Tab.~{\ref{#1}}}

\newcommand{\ov}{\overline}

\newcommand{\LM}{\mathrm{LM}}
\newcommand{\RD}{\mathrm{RD}}
\newcommand{\Full}{\mathrm{Full}}

\begin{document}
\date{\today}

\title{Chiral symmetry breaking and the generation of light hadron masses}
\author{M.~Denissenya}
\email{mikhail.denissenya@uni-graz.at}
\affiliation{Institut f\"ur Physik, FB Theoretische Physik, Universit\"at
Graz, A--8010 Graz, Austria}
\author{L.~Ya.~Glozman}
\email{leonid.glozman@uni-graz.at}
\affiliation{Institut f\"ur Physik, FB Theoretische Physik, Universit\"at
Graz, A--8010 Graz, Austria}

\begin{abstract}
A quantitative study of the effects of dynamical chiral symmetry breaking on the mass generation of the low-lying hadrons in a dynamical lattice QCD simulation is presented. The evolution of light hadron masses upon increasing the number of low-lying Dirac operator eigenmodes in the quark propagators is confronted with the hadron masses obtained upon removal of such low-lying eigenmodes from standard full quark propagators. The low-lying chiral symmetry breaking modes provide roughly two third of the nucleon and $\rho$ masses, while the $a_1$ mass is affected to a much smaller degree.
\end{abstract}
\maketitle

\section{Introduction}

It has been recognized for a long time that the approximate chiral symmetry
and its dynamical breaking in the QCD vacuum represents an important
ingredient for hadron physics in the light quark sector. The mass of such hadrons
originates almost entirely from the energy of the quantized gluonic
field. These gluonic interactions lead to both, dynamical chiral symmetry breaking (D$\chi$SB) and confinement. Apriori it is not clear which of these
two most important phenomena or both are of importance for the generation of mass of light hadrons. It is well established that the exceptionally low mass of the pions
is a consequence of D$\chi$SB. According to some model views the
nucleon and $\rho$-meson masses are as well due to the physics of D$\chi$SB and are
not related to confinement. In this Brief Report we aim to investigate this
issue by quantifying the effects of the quark condensate of the vacuum on
the mass generation of the low-lying hadrons in a dynamical lattice simulation.

Several qualitative lattice studies addressing this question have been performed in the past
\cite{Neff:2001zr,DeGrand:2003sf,Bali:2010se}. One of the conclusions was 
that the pion mass is saturated by the D$\chi$SB physics, in agreement with
theoretical expectations. The current study follows our previous work
\cite{Lang:2011vw,Glozman:2012fj} where we
investigated the effects of artificially restoring the chiral symmetry
by  removing the low-lying modes from the valence quark propagators. Indeed,
according to the Banks--Casher relation \cite{Banks:1979yr} a density of
the quasi-zero modes of the Dirac operator represents the quark condensate
of the vacuum. It has turned out that all low-lying hadrons, except for the
pions, survive this artificial unbreaking of the chiral symmetry. Chiral
symmetry appears restored while the nucleon and rho masses do not drop down. At
some truncation level we reached a regime where the hadron mass was entirely 
chirally symmetric and consequently its generation is not related to D$\chi$SB.
In this world all hadrons arrange themselves into  multiplets of the chiral
group. This does not tell, however, that in the real world D$\chi$SB is
unimportant for the masses of the low-lying hadrons. Indeed, the size of the
respective splittings, e.g. $a_1 - \rho$, is of the order of the hadron mass.
Consequently we should expect at least a sizeable contribution of D$\chi$SB dynamics
on the low-lying hadron mass generation.

In this study we want to gain deeper inside in the role of D$\chi$SB for the generation of individual light hadron's masses. Therefore we consider not only Dirac low mode reduced quark propagators, as in previous works, but also explore hadrons constructed out of the low mode part exclusively.
This will allow for a quantitative comparison of the contribution of the low-lying D$\chi$SB Dirac modes to the mass generation of different light hadrons.

\section{Our strategy}

\subsection{Low mode only vs. low mode reduced propagators}
 The quark propagator in a given gauge background 
can be decomposed into a sum of the eigenmodes
 of the Dirac operator. According to the Banks--Casher relation, a density
 of the lowest quasi-zero eigenmodes is related to the quark
 condensate of the vacuum. Consequently, the low-lying modes
 represent the physics of D$\chi$SB. The propagator built
 from the $k$ lowest-lying modes (by magnitude) only, we call low mode (LM) propagator,
\begin{figure*}[t]
  \centering
  \hspace*{12pt}
  \includegraphics[width=0.45\textwidth]{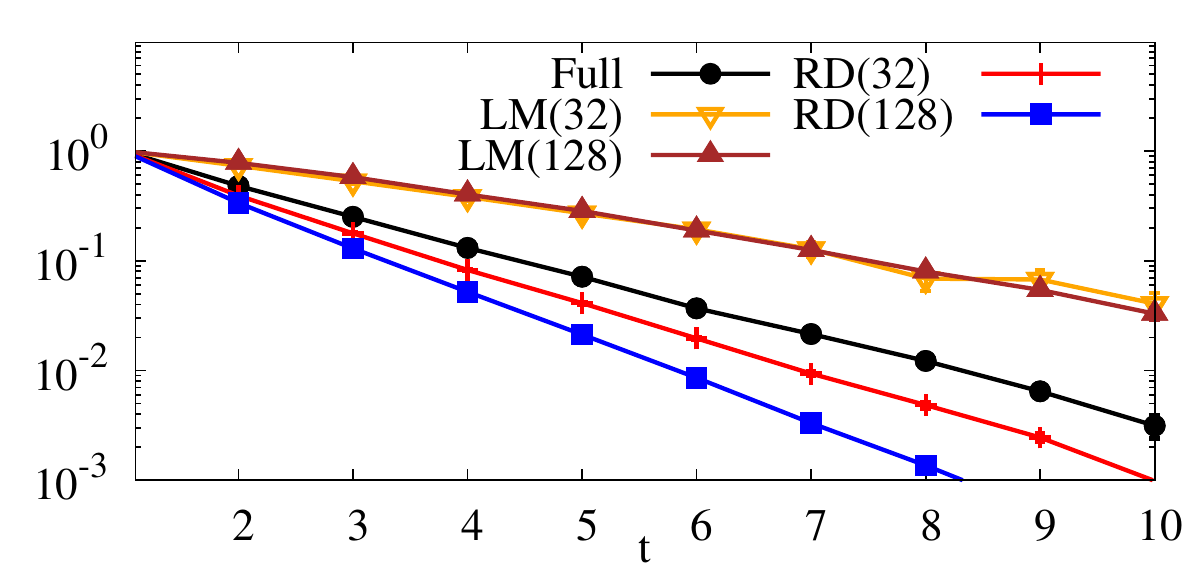}\hfill
  \includegraphics[width=0.45\textwidth]{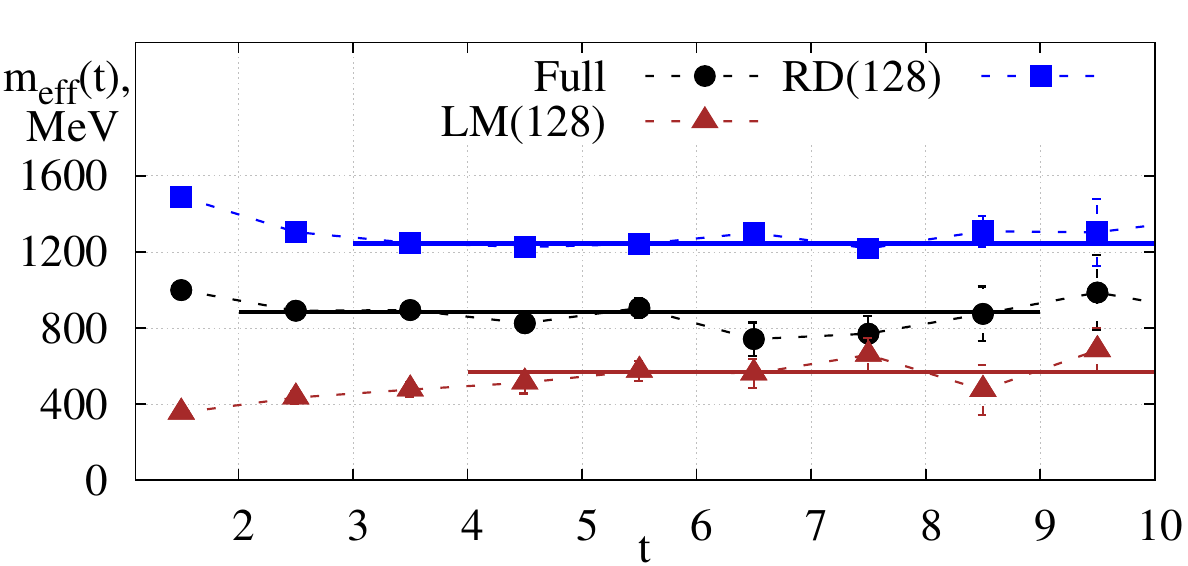}
  \hspace*{24pt}\hfil\\
  \hspace*{12pt}
  \includegraphics[width=0.45\textwidth]{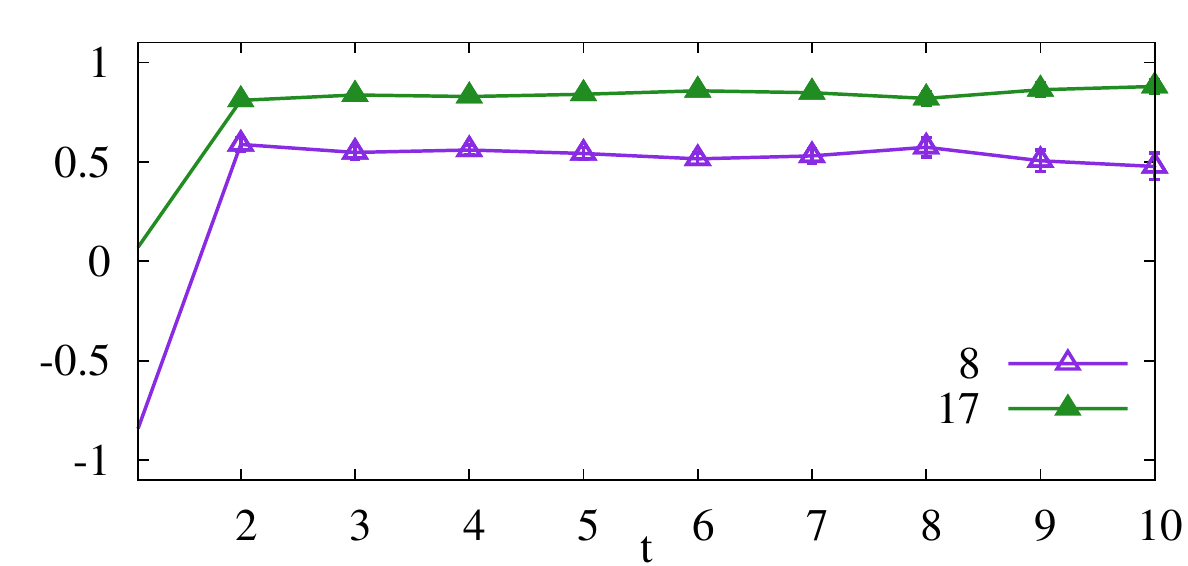}\hfill
  \includegraphics[width=0.45\textwidth]{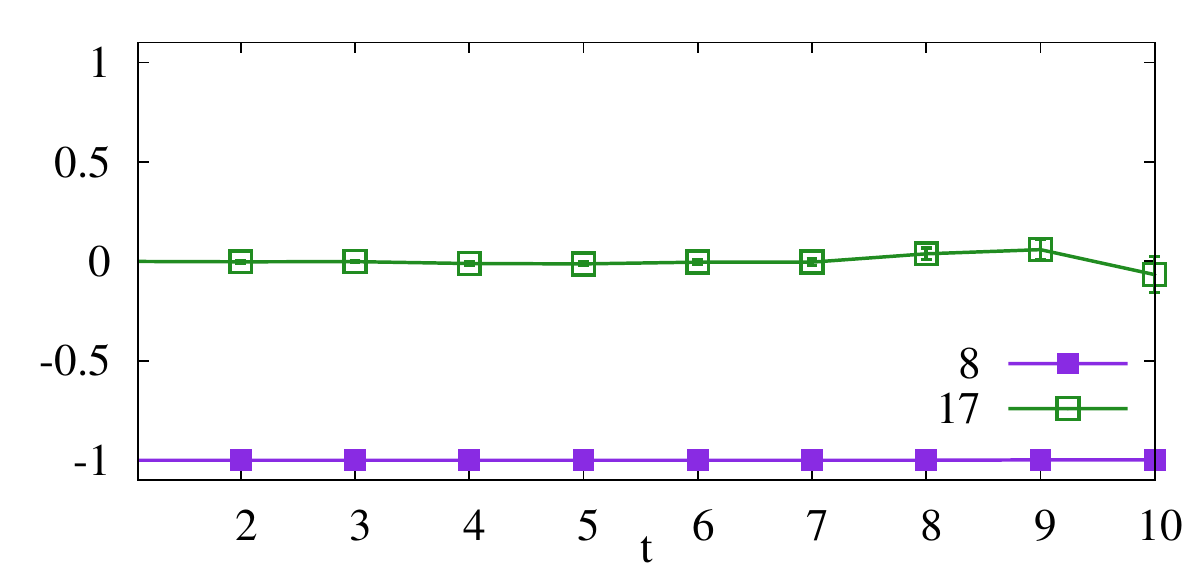}
  \hspace*{24pt}\hfil\\
  \caption{The $\rho$-meson ($J^{PC}=1^{--}$) from different truncation stages of low mode (LM) and reduced (RD) quark propagators.
  The eigenvalues of the correlation matrix are shown in the upper left plot,
  the effective masses in the upper right.
  The bottom row shows the eigenvectors of the variational method of LM(128) (left) and  RD(128) (right). 
}\label{fig:rho}
\end{figure*}

 \begin{equation}\label{eq:LM}
  S_{\LM(k)}=\sum_{i=1}^{k}\,\frac{1}{\lambda_i}\,|\lambda_i\rangle \langle \lambda_i| \gamma_5\;, 
 \end{equation}

\noindent
where $\lambda_i$ and $|\lambda_i\rangle$ are the real eigenvalues and
the corresponding eigenvectors of the Hermitian Dirac operator $\gamma_5 D$.
 A propagator constructed by subtracting the  lowest-lying modes from the 
 complete set, we call a reduced (RD) propagator,
 \begin{equation}\label{eq:RD}
  S_{\RD(k)}=S_{\Full}-S_{\LM(k)}\;. 
 \end{equation}
 
 \noindent
 Such a propagator does not incorporate the chiral symmetry breaking physics,
 but takes into account (at least most of) the confining properties \cite{Lang:2011vw,Glozman:2012fj}.
 
 From the low mode and reduced quark propagators at a given truncation level $k$
 we construct hadron propagators and study the corresponding correlation functions. When we observe an exponential decay of the latter, we interpret it as a signal of the state and extract
 its mass. By comparing hadron masses from both, low mode and reduced
 propagators at different $k$, we can judge to what extent D$\chi$SB 
 is important for the mass generation of the respective hadron.
 We choose truncation levels of $k = 32, 64$ and $128$ corresponding to truncation
 energies of 42, 71 and 117 MeV, respectively.
 
 For each hadron we use a set of different interpolators and analyse the
 corresponding cross-correlation matrix with the variational method
 by solving the generalized eigenvalue problem \cite{Luscher:1990ck,Michael:1985ne}. The quantum channels and corresponding interpolators 
 are listed in \tab{tab:ints} where the enumeration of the interpolators has been inherited from
\cite{Engel:2011aa,Engel:2013ig}. 
Notations $n$, $w$ and $\partial_i$ stand for the Jacobi smeared narrow, wide and derivative sources, respectively \cite{Gattringer:2008vj}.

  \begin{table}[h]
    \begin{center}
    \begin{tabular}{|c|c|}
    \hline
    \hline
        $\#$& $\rho$\\
    \hline      
        8   & $\ov{q}_w \gamma_k\gamma_t q_w$  \\
        17  & $\ov{q}_{\partial_i} \gamma_k q_{\partial_i}$\\
     \hline
     \hline    
        $\#$& $a_1$ \\
     \hline      
        1   & $\ov{q}_n \gamma_k \gamma_5 q_n$    \\
        4   & $\ov{q}_w \gamma_k \gamma_5 q_w$  \\
     \hline
     \hline      
       $\#$ & $N^{+}$\\
     \hline     
       1    &$\varepsilon_{abc} q_a (q^T_bC\gamma_5 q_c)$  , $n(nn)$\\     
       18   & $\varepsilon_{abc} i q_a (q^T_bC\gamma_t\gamma_5 q_c)$, $w(ww)$ \\

    \hline
    \end{tabular}
    \end{center}
    \caption{Interpolators for the
             $\rho(1^{--})$,  $a_1(1^{++})$ mesons and the nucleon $N^+$.
             }\label{tab:ints}
  \end{table}
  

  \begin{figure*}[t!]
  \hspace*{12pt}
  \includegraphics[width=0.45\textwidth]{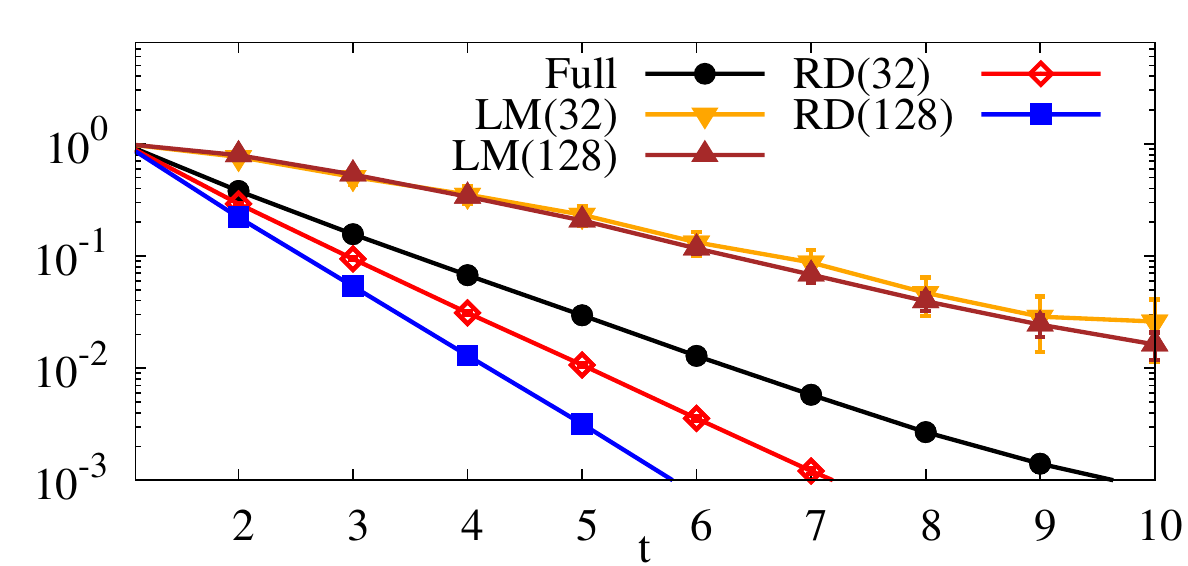}\hfill
  \includegraphics[width=0.45\textwidth]{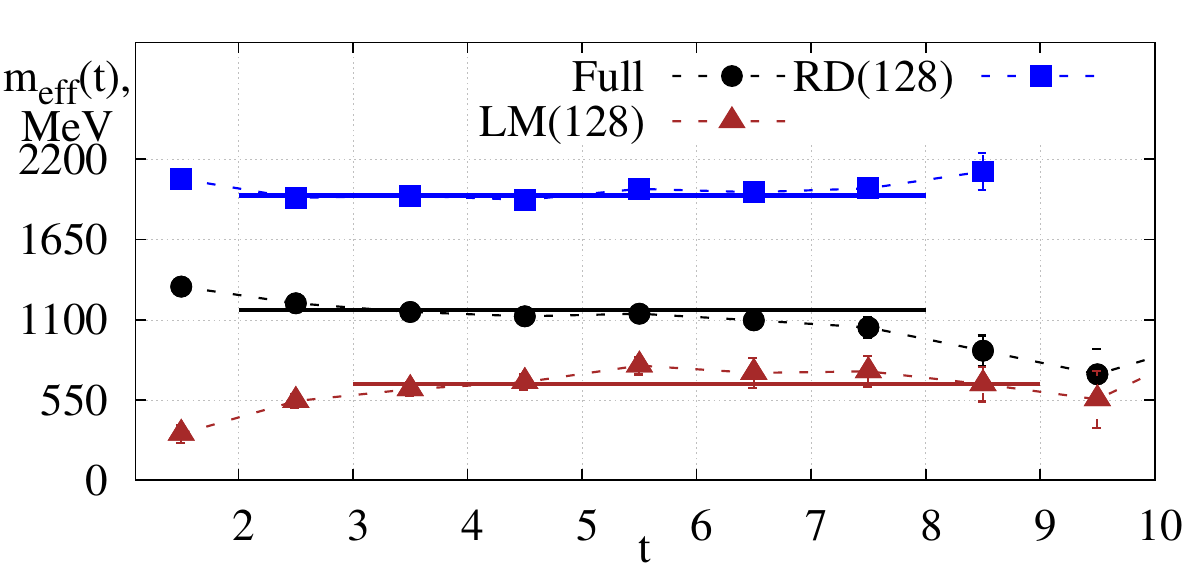}
  \hspace*{24pt}\hfil\\
  \hspace*{12pt}
  \includegraphics[width=0.45\textwidth]{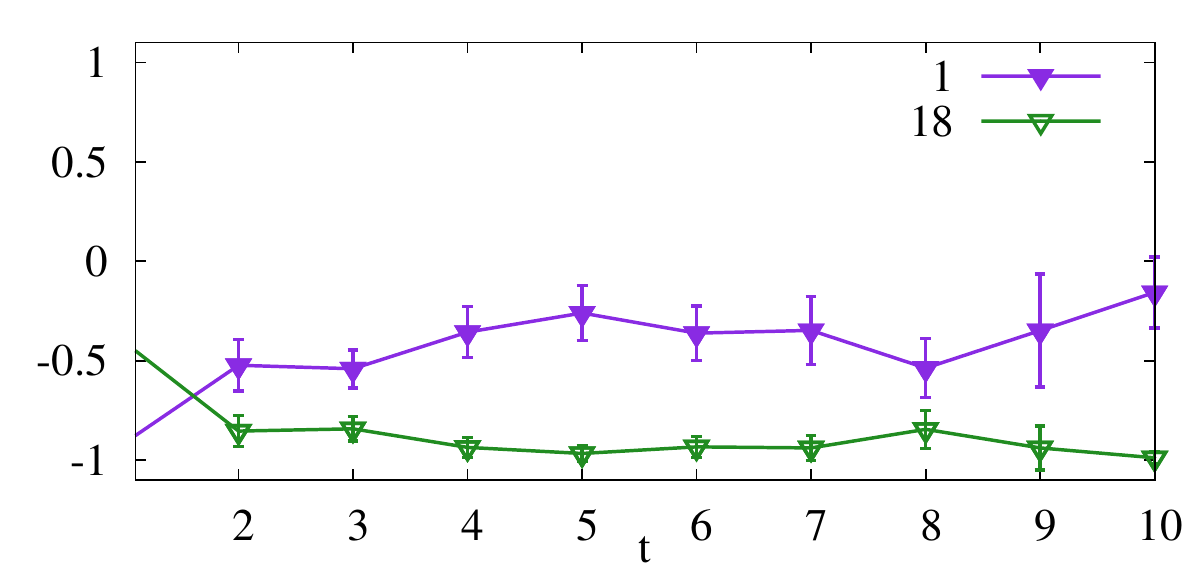}\hfill
  \includegraphics[width=0.45\textwidth]{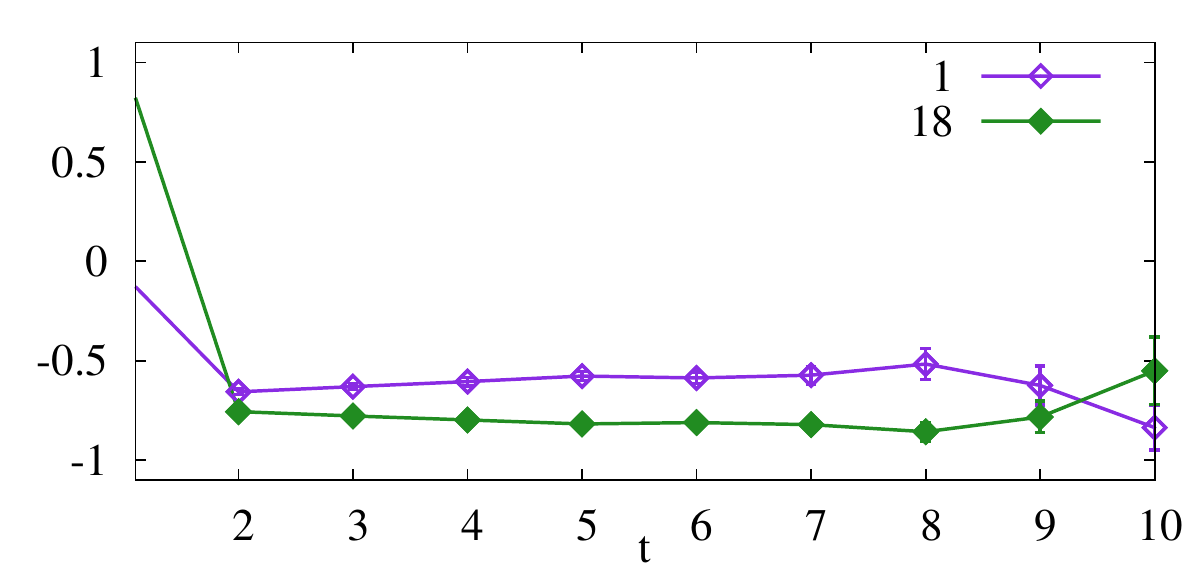}
  \hspace*{24pt}\hfil\\
  \caption{The nucleon $N$ ($J^{P}=1/2^+$) from different truncation stages of low mode (LM) and reduced (RD) quark propagators.
  The eigenvalues of the correlation matrix are shown in the upper left plot,
  the effective masses in the upper right.
  The bottom row shows the eigenvectors of the variational method of LM(32) (left) and  RD(32) (right).
   }\label{fig:nuc} 
  \end{figure*}
  
  \begin{figure*}[t]

  \hspace*{12pt}
  \includegraphics[width=0.45\textwidth]{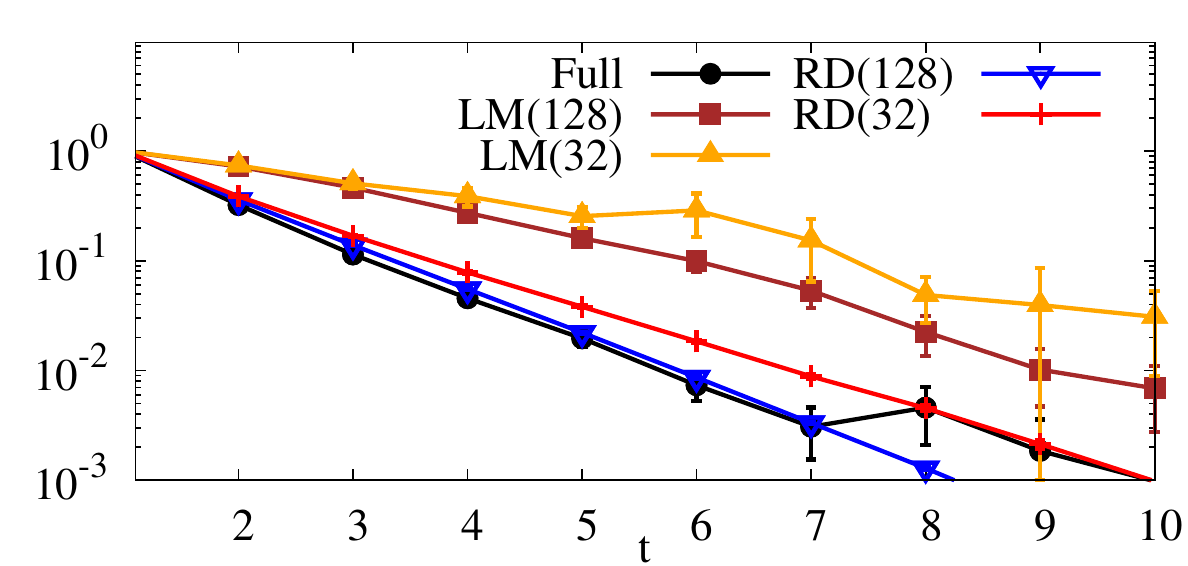}\hfill
  \includegraphics[width=0.45\textwidth]{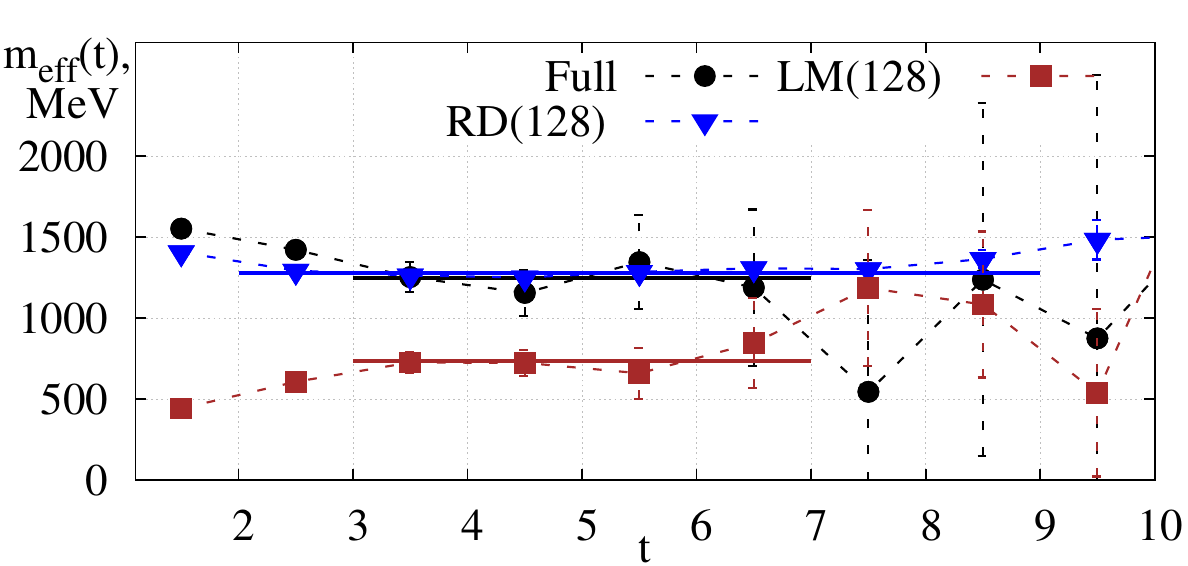}
  \hspace*{24pt}\hfil\\
  \hspace*{12pt}
  \includegraphics[width=0.45\textwidth]{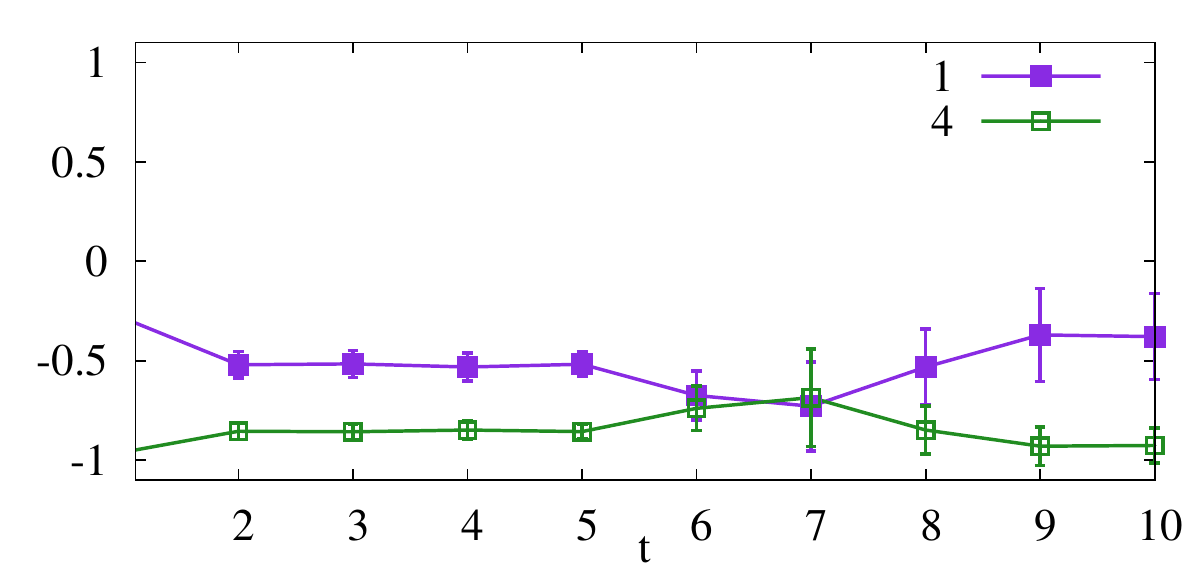}\hfill
  \includegraphics[width=0.45\textwidth]{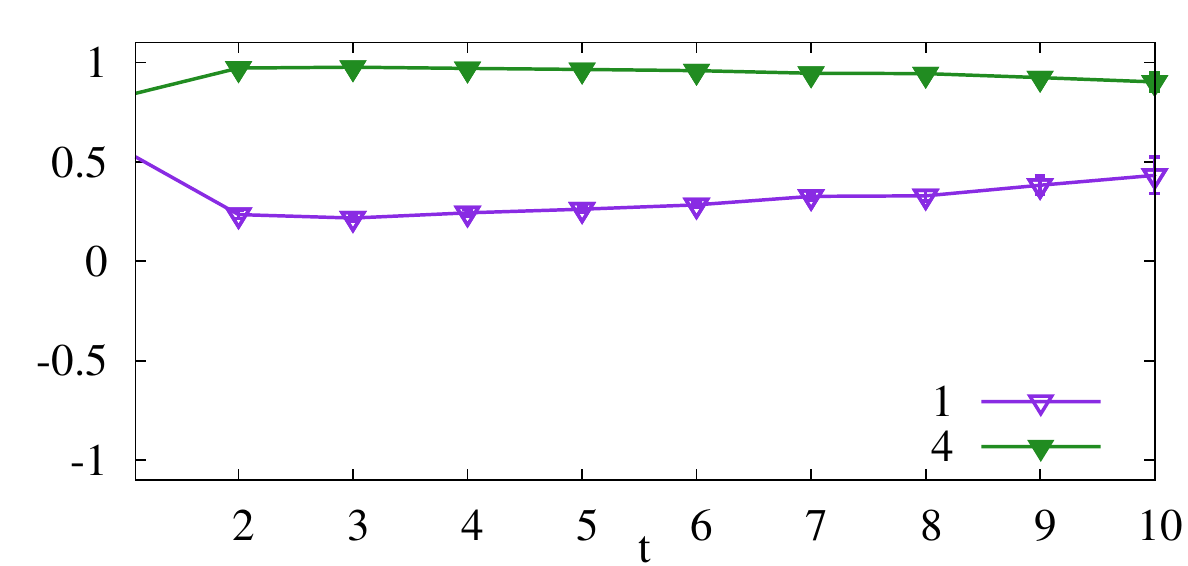}
  \hspace*{24pt}\hfil\\
  \caption{
  The $a_1$-meson ($J^{PC}=1^{++}$) from different truncation stages of low mode (LM) and reduced (RD) quark propagators.
  The eigenvalues of the correlation matrix are shown in the upper left plot,
  the effective masses in the upper right.
  The bottom row shows the eigenvectors of the variational method of LM(128) (left) and  RD(128) (right).
  }\label{fig:a1}
  \end{figure*}


  \begin{figure*}[ht!]
  \centering
    \hspace*{12pt}
    \includegraphics[width=0.55\textwidth]{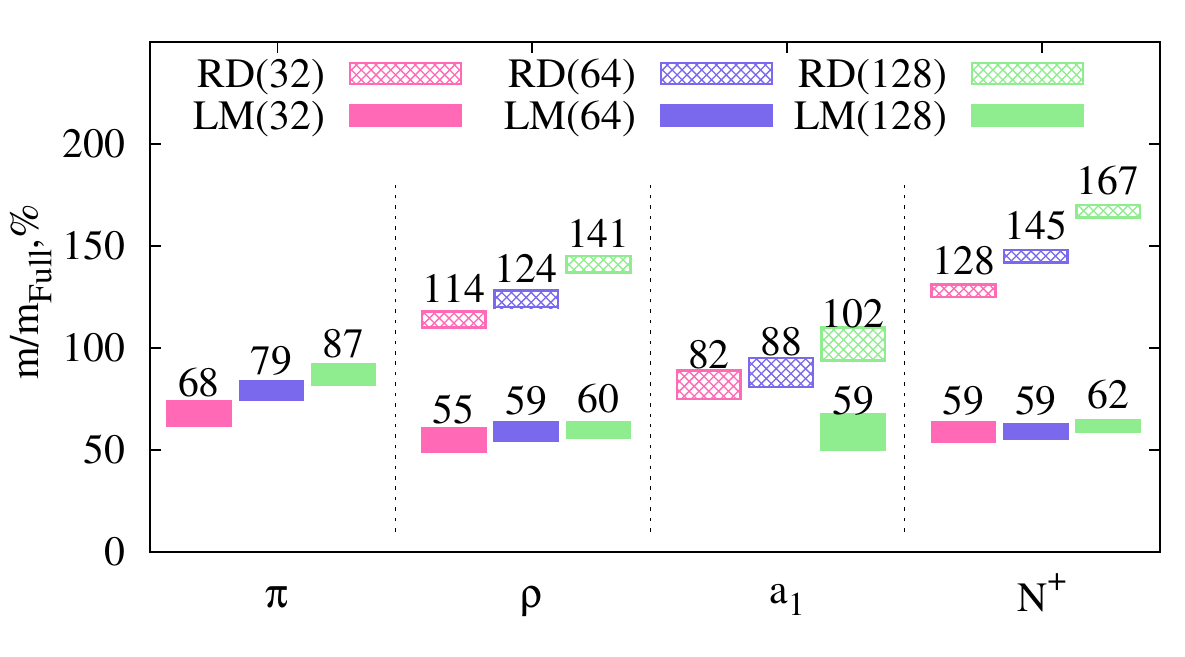}
    \hspace*{12pt}\hfil\\
    \hspace*{12pt}
    \caption{Relative masses of hadrons from low mode quark propagators and reduced quark propagators.}\label{fig:histo}
  \end{figure*}

\subsection{Lattice setup}
  We adopt 160 gaugefield configurations generated with $n_f=2$ dynamical chirally improved
fermions \cite{Gattringer:2000js} on a $16^3\times 32$ lattice with lattice spacing
$a=\unit[0.144(1)]{fm}$. The corresponding pion mass is $M_{\pi}=322(5)$ MeV in this ensemble \cite{Gattringer:2008vj}.

\section{Evolution of the hadron masses of low mode vs. reduced propagators}
 
 \subsection{$\pi$} 
  The pion mass is saturated by the low-lying modes only \cite{DeGrand:2003sf}.
  At   $k=128$ the  pion mass is reproduced within the error bars. 
  On the other hand, when one removes
 the same number of low modes from the complete set one artificially removes the pion from the spectrum in the sense of a manifest loss of the exponential decay of the correlator function \cite{Lang:2011vw}.   
   
\subsection{$\rho$ and $N$}
Figs.~\ref{fig:rho}  and \ref{fig:nuc} show the correlation matrix eigenvalues, effective mass plots and the eigenvectors from the variational method of the $\rho$-meson ($J^{PC}=1^{--}$) and the nucleon $N$ ($J^P=1/2^+$), respectively, using low mode vs. reduced quark propagators. 

A clear signal of a bound state using the $\LM(32)$ quark propagator 
 in both channels is visible with the masses roughly two third of the actual $\rho$ and $N$  masses.
When further increasing the number of included modes in the LM propagators
the masses increase very slowly so that the masses extracted with LM(32) and LM(128) practically coincide. This tells that the 32 lowest-lying, chiral symmetry breaking
  modes provide a bulk (two third) of the rho and nucleon masses and
  the remaining one third  is due to higher-lying and high-lying eigenmodes which are not related to D$\chi$SB. 
  
If, in contrast, we remove the low-lying modes from the full propagators, i.e., we adopt RD propagators, both, the $\rho$ and $N$, survive. Their mass is chirally symmetric, e.g. 
  the $\rho$-meson gets
  degenerate with its chiral partner $a_1$. The masses of these chirally symmetric hadrons are higher than the masses obtained with the untruncated full propagators. 
  Note that there is no additive law for hadron masses, so that the sum of the
  LM mass and of the RD mass does not coincide with the full mass. The physics of the mass generation
  for the $\rho$ and $N$ obtained from the chiral symmetry breaking low-lying modes and
  from the chiral-invariant higher modes is very different.

  \subsection{$a_1$}
  For the $a_1$-meson the effect of the low-lying modes is essentially less important, see \fig{fig:a1}. From the lowest 32 modes we do not see clear exponential decay, i.e. a bound state is absent. In contrast, a clear state is seen if we remove the same amount of the low-lying modes from the quark propagators. This tells that it is the confining higher-lying and high-lying modes but not the low-lying chiral symmetry breaking modes that are the most essential for the mass generation of the $a_1$ meson.


\section{Conclusions}
The evolution of the light hadron masses upon increasing  the number of the included low-lying Dirac eigenmodes in the LM  quark propagators
and upon increasing the number of the removed low modes in the RD propagators, Eqs.~(\ref{eq:LM}) and (\ref{eq:RD}), is summarized in \fig{fig:histo}.
We conclude that -- unlike the mass of the pion -- the
masses of $\rho$ and $N$ grow very slowly with the number of included low modes in the LM propagators above $k=32$,
  see  \fig{fig:histo} . These low modes provide a large contribution to their masses
  and give rise up to two third of the corresponding full mass. To fill the remaining
mass for the $\rho$-meson and the nucleon one needs to take 
  into account the high-lying Dirac eigenmodes.  In contrast, it is the higher-lying modes that are the
  most   essential for the mass of the $a_1$-meson. 

\begin{acknowledgments}
 We thank Christian Lang and Mario ~Schr\"ock for collaboration and many fruitful discussions.
  Support from the Austrian Science Fund (FWF) through the grants DK W1203-N16, P21970-N16 and
P26627-N16  is acknowledged.
\end{acknowledgments}

\bibliographystyle{apsrev4-1}
\bibliography{msh} 

\end{document}